\documentclass{article}

  \usepackage[preprint]{neurips_2026}
  \makeatletter\renewcommand{\@notice}{}\makeatother

\usepackage[utf8]{inputenc}
\usepackage[T1]{fontenc}
\usepackage{hyperref}
\hypersetup{hidelinks}
\usepackage{url}
\usepackage{booktabs}
\usepackage{array}
\newcolumntype{L}[1]{>{\raggedright\arraybackslash}p{#1}}
\usepackage{amsfonts}
\usepackage{nicefrac}
\usepackage{microtype}
\usepackage{graphicx}
\usepackage{xcolor}
\usepackage{pgfplots}
\pgfplotsset{compat=1.18}
\usepackage{caption}
\usepackage{enumitem}
\setlist{itemsep=1pt, topsep=3pt, parsep=0pt}

\newcommand{\baselinecurveA}{(0.0,0.0) (2.31,0.0) (4.62,0.0) (6.94,0.0) (9.25,0.0) (11.56,0.0) (13.87,0.0) (16.18,0.0) (18.5,0.0) (20.81,0.0) (23.12,0.0) (25.43,0.0) (27.75,0.0) (30.06,0.0) (32.37,0.0) (34.68,0.0) (36.99,0.0) (39.31,0.0) (41.62,0.0) (43.93,0.0) (46.24,0.0) (48.55,0.0) (50.87,0.0) (53.18,0.0) (55.49,0.0) (57.8,0.0) (58.38,0.1) (58.96,0.2) (59.54,0.4) (60.12,0.5) (60.69,0.5) (61.27,0.6) (61.85,0.7) (62.43,0.7) (63.01,0.7) (63.58,0.8) (64.16,1.0) (64.74,1.1) (65.32,1.1) (65.9,1.3) (66.47,1.3) (67.05,1.5) (67.63,1.8) (68.21,2.0) (68.79,2.1) (69.36,2.5) (69.94,2.6) (70.52,2.6) (71.1,2.9) (71.68,2.9) (72.25,2.9) (72.83,3.0) (73.41,3.0) (73.99,3.2) (74.57,3.2) (75.14,3.6) (75.72,3.6) (76.3,3.9) (76.88,4.2) (77.46,4.5) (78.03,4.8) (78.61,5.0) (79.19,5.0) (79.77,5.1) (80.35,5.5) (80.92,5.9) (81.5,6.2) (82.08,6.3) (82.66,6.5) (83.24,7.9) (83.82,8.5) (84.39,9.5) (84.97,10.7) (85.55,11.0) (86.13,11.2) (86.71,11.8) (87.28,12.2) (87.86,15.9) (88.44,18.5) (89.02,19.0) (89.6,21.6) (90.17,22.0) (90.75,23.0) (91.33,27.0) (91.91,28.3) (92.49,31.6) (93.06,34.4) (93.64,38.0) (94.22,39.3) (94.8,44.1) (95.38,45.8) (95.95,54.5) (96.53,56.7) (97.11,58.3) (97.69,65.5) (98.27,66.2) (98.84,66.9) (99.42,70.3) (100.0,97.9)}
\newcommand{\baselinemarkA}{(87,14.4)}

\newcommand{\baselinepctA}{87}
\newcommand{\baselinecurveB}{(0.0,0.0) (4.6,0.0) (9.2,0.0) (13.79,0.0) (18.39,0.0) (22.99,0.0) (27.59,0.0) (32.18,0.0) (36.78,0.0) (41.38,0.0) (45.98,0.0) (50.57,0.0) (55.17,0.0) (59.77,0.0) (64.37,0.0) (67.82,0.4) (68.97,0.5) (70.11,0.6) (71.26,2.3) (72.41,2.6) (73.56,2.7) (74.71,2.7) (75.86,3.0) (77.01,3.1) (78.16,3.1) (79.31,3.8) (80.46,4.6) (81.61,4.8) (82.76,4.8) (83.91,7.3) (85.06,7.7) (86.21,8.3) (87.36,8.6) (88.51,9.3) (89.66,10.3) (90.8,11.5) (91.95,11.6) (93.1,20.0) (94.25,30.3) (95.4,34.0) (96.55,53.8) (97.7,63.8) (98.85,69.6) (100.0,94.3)}
\newcommand{\baselinemarkB}{(99,78.2)}

\newcommand{\baselinepctB}{99}

\newcommand{\survaddA}{(1,8) (2,8) (3,7) (4,42) (5,33) (6,1) (7,1) (8,2) (9,11) (10,2) (11,1) (12,2) (13,5) (14,9) (15,6) (16,2) (17,2) (18,10) (19,17) (20,16) (21,15) (22,1) (23,0) (24,17) (25,151) (26,401) (27,1) (28,100) (29,9) (30,106) (31,4) (32,46) (33,0) (34,14) (35,77) (36,68) (37,1) (38,0) (39,853) (40,38) (41,153)}
\newcommand{\survaliveA}{(1,4) (2,1) (3,4) (4,15) (5,13) (6,0) (7,0) (8,0) (9,4) (10,0) (11,0) (12,0) (13,2) (14,4) (15,0) (16,0) (17,0) (18,2) (19,4) (20,12) (21,6) (22,0) (23,0) (24,6) (25,112) (26,401) (27,0) (28,100) (29,2) (30,60) (31,0) (32,29) (33,0) (34,6) (35,51) (36,47) (37,1) (38,0) (39,853) (40,23) (41,153)}
\newcommand{\survmaxA}{853}
\newcommand{\survnA}{41}
\newcommand{\survpctA}{85.5}
\newcommand{\survzeroA}{15}
\newcommand{\survhalfA}{3}
\newcommand{\survhalfpctA}{7.3}
\newcommand{\survprenA}{26}

\newcommand{\survprepctA}{35.8}

\newcommand{\survpathfirstA}{2026-04-06}
\newcommand{\survfirstA}{2025-09-01}

\newcommand{\survvisiblefmtA}{1{,}915}
\newcommand{\survprefmtA}{686}
\newcommand{\survaddB}{(1,3) (2,19) (3,32) (4,18) (5,81) (6,47) (7,17) (8,24) (9,59) (10,45) (11,67) (12,49) (13,50) (14,26) (15,107) (16,5) (17,5) (18,82) (19,3) (20,62) (21,2) (22,151) (23,13) (24,98) (25,25) (26,55) (27,1) (28,2) (29,8) (30,2) (31,18) (32,14) (33,86) (34,7) (35,20) (36,29) (37,61) (38,228) (39,1) (40,10) (41,71)}
\newcommand{\survaliveB}{(1,0) (2,2) (3,2) (4,0) (5,5) (6,0) (7,1) (8,1) (9,0) (10,0) (11,0) (12,2) (13,7) (14,0) (15,0) (16,0) (17,0) (18,1) (19,0) (20,0) (21,0) (22,1) (23,0) (24,4) (25,2) (26,53) (27,1) (28,0) (29,8) (30,2) (31,0) (32,11) (33,64) (34,0) (35,20) (36,29) (37,0) (38,37) (39,1) (40,10) (41,71)}

\newcommand{\survnB}{41}
\newcommand{\survpctB}{19.7}
\newcommand{\survzeroB}{18}
\newcommand{\survhalfB}{7}
\newcommand{\survhalfpctB}{17.1}

\newcommand{\survaddedfmtB}{1{,}703}

\newcommand{\monthlabelsA}{{Jun\\2025},Jul,Aug,Sep,Oct,Nov,Dec,{Feb\\2026},Mar,Apr,May,Jun,Jul,Aug}

\newcommand{\monthcoordsA}{2025-06,2025-07,2025-08,2025-09,2025-10,2025-11,2025-12,2026-02,2026-03,2026-04,2026-05,2026-06,2026-07,2026-08}
\newcommand{\monthdiscA}{(2025-06,0) (2025-07,0) (2025-08,0) (2025-09,3) (2025-10,0) (2025-11,0) (2025-12,0) (2026-02,14) (2026-03,11) (2026-04,31) (2026-05,5) (2026-06,0) (2026-07,2) (2026-08,0)}
\newcommand{\monthrestA}{(2025-06,6) (2025-07,9) (2025-08,19) (2025-09,35) (2025-10,1) (2025-11,9) (2025-12,29) (2026-02,76) (2026-03,60) (2026-04,149) (2026-05,3) (2026-06,1) (2026-07,2) (2026-08,1)}

\newcommand{\identlabelsA}{Claude Opus 4.8,Claude,Claude Opus 4.6,Claude Opus 4.7 (1M context),Claude Opus 4.6 (1M context),Claude Sonnet 4.6}
\newcommand{\identcountsA}{(2,1) (3,2) (5,3) (14,4) (17,5) (25,6)}
\newcommand{\identnA}{6}
\newcommand{\identticksA}{1,2,3,4,5,6}

\newcommand{\identtotalA}{66}
\newcommand{\identbareA}{3}
\newcommand{\keynA}{3}
\newcommand{\keyunionA}{121}
\newcommand{\keyunionpctA}{26.0}
\newcommand{\keymissedA}{55}
\newcommand{\keymissedpctA}{45.5}

\newcommand{\keyerafactorA}{12.3}
\newcommand{\keytraceA}{8}

\newcommand{\keysecondA}{55}
\newcommand{\keythirdA}{2}
\newcommand{\sweeptopwordA}{cursor}
\newcommand{\sweeptopcountA}{58}
\newcommand{\kwrateA}{40.8}
\newcommand{\kwhitsA}{190}
\newcommand{\kwnotrailerA}{124}

\newcommand{\qsinglerevA}{7}

\newcommand{\qsingleshareA}{10.3}
\newcommand{\qsinglefirstA}{2026-04-17}
\newcommand{\qunionrevA}{15}

\newcommand{\qunionshareA}{14.4}
\newcommand{\qunionfirstA}{2026-04-06}
\newcommand{\qfollowrevA}{41}

\newcommand{\qfollowshareA}{14.5}
\newcommand{\qfollowfirstA}{2025-09-01}
\newcommand{\qspreadA}{4.2}
\newcommand{\qratioA}{5.9}
\newcommand{\qdaysA}{228}
\newcommand{\qsinglerevB}{39}

\newcommand{\qsingleshareB}{78.2}
\newcommand{\qsinglefirstB}{2025-10-02}
\newcommand{\qunionrevB}{39}

\newcommand{\qunionshareB}{78.2}
\newcommand{\qunionfirstB}{2025-10-02}
\newcommand{\qfollowrevB}{41}

\newcommand{\qfollowshareB}{80.3}
\newcommand{\qfollowfirstB}{2025-10-02}
\newcommand{\qspreadB}{2.1}

\title{What Git Does Not Record:\\ Provenance Legibility in Two AI-Assisted Research Repositories%
  \thanks{Accepted at the NeurIPS 2026 Workshop on AI-Native Academia:
  Authorship, Peer Review, and Conference Governance under AI.}}

\workshoptitle{AI-Native Academia}

\hypersetup{
  pdftitle={What Git Does Not Record: Provenance Legibility in Two AI-Assisted Research Repositories},
  pdfauthor={Koutian Wu, Zesen Huang, and Hejia Geng}
}
\author{%
  Koutian Wu \\
  The University of Texas at Austin \\
  Corresponding author: \href{mailto:ktwu@utexas.edu}{\texttt{ktwu@utexas.edu}} \\
  \And
  Zesen Huang \\
  University of California, Los Angeles \\
  \And
  Hejia Geng \\
  University of Oxford \\
}

\begin{document}

\maketitle

\begin{abstract}
Publishers and funders increasingly ask authors to disclose how AI tools contributed to a
piece of research. In computational work the repository is where an author or an auditor
would look for the answer, and we ask what it can actually supply. We audit two single-investigator repositories by the same author in the same field, each carrying a different
class of agent residue: a \emph{prescriptive} trace of standing rule files and a
\emph{narrative} trace of a running agent scratchpad, in repositories of $466$ and $103$
commits. The two are
legible in opposite ways. $85.6\%$ of the lines ever committed to the prescriptive trace are
visible at the current revision, and a standing rule states a conclusion and discards the
episode behind it; the narrative trace records episodes and shows $21.8\%$. Counted by
revision the gap is far smaller: $15$ of the prescriptive trace's $41$ revisions leave no
visible line, against $18$ of the narrative trace's $41$. The \texttt{Co-Authored-By} trailer, the field
an automated audit reads, is one of three AI trailer keys Project~A carries and covers $66$ of the
$121$ commits carrying any. In Project~B it has low sensitivity against an explicit
within-repository proxy: $39$ commits revise a file maintained to hold an agent's working notes and $1$
($2.6\%$) carries the trailer. Project~A's repository-level rate is unstable in time too,
stepping from $2.7\%$ to $17.7\%$ across a boundary we did not specify in advance. Repository
channels disagree, and neither rate can be read as an involvement rate; we do not isolate the
cause, and say so. We give a field-by-field account of which provenance claims such traces
support, which they cannot, and which they answer confidently and wrongly, construct an
auditable evidence ledger, and derive the requirements this places on any system meant to
record research provenance as it happens.
\end{abstract}

\section{Introduction}

The disclosure requirements now attached to AI use in research ask authors to describe
something that happened months earlier. Journals require a statement of how generative tools
contributed \citep{nature2023tools}; conferences ask for the role LLMs played; reviewers,
when suspicious, ask for evidence. These policies specify where the statement goes, not
where its content is to come from. In computational research there is one obvious
candidate: the project's version-control repository is already kept, timestamped, shared,
and treated as the authoritative record of how the work was built, and it is the first
place an author reconstructing their own process, or an auditor checking it, would look.
Whether it can carry that weight is an empirical question. A repository records a narrow
thing: a sequence of snapshots of a file tree, each annotated with a name, an email
address, two timestamps, and a free-text message, all supplied by the committer and none
verified. Projects using agentic coding tools commonly add one thing more, a persistent
artifact holding the agent's standing instructions or its working notes between sessions.
Where such an artifact is tracked, it is the closest thing to a contemporaneous record of
AI involvement an ordinary repository contains, and it is what an auditor would read. The
two repositories we audit carry structurally different artifacts of this kind, and what
each presents to a reader is the subject of the paper. It is a comparison of two cases,
not a controlled manipulation, and we claim only what that licenses.

\paragraph{Design.}
We compare two repositories from a single investigator working in the environmental sciences
(Table~\ref{tab:arms}). Project~A ($466$ commits) was developed with an agentic coding
assistant whose persistent artifact is a directory of standing rule files: instructions the
agent is to follow on future work. We call this a \emph{prescriptive} trace. Project~B ($103$
commits) is a Cursor-assisted project, evidenced by a versioned agent scratchpad tracked
across $39$ commits; its artifact is a single running document of plans, findings and status,
rewritten as work proceeds. We call this a \emph{narrative} trace. Author, discipline and
version-control substrate are shared; subject, period, commit volume, research phase and
how heavily each tool was used are not, and we manipulated none of them. Tool, artifact
class, project and time are confounded, so no difference we report can be attributed to the
tool alone. What the shared author buys is narrower and still worth having: it removes the
most obvious rival explanation for a difference in record-keeping, and it makes the
within-repository comparisons in Section~\ref{sec:disclosure} comparisons within one
person's practice.

\paragraph{Findings.} The two artifacts are legible in opposite ways, and the failures are
complementary. The prescriptive trace accumulates, keeping $85.6\%$ of its committed lines
at the current revision, but a standing rule records a conclusion rather than the episode
that produced it. The narrative trace records episodes and shows only $21.8\%$: the rest
is in the object database, recoverable by walking the file's history rather than by reading
its contents, which is the distinction that matters for an audit. So the artifact that
records the process does not show it, and the artifact that shows itself does not record
the process. Both figures presuppose the artifact was committed at all, a choice neither
tool prompts for and neither warns against. The commit log fares worse than either. The field
an automated audit reads, the \texttt{Co-Authored-By} trailer, is absent from $38$ of the
$39$ Project~B commits that revise the agent's own scratchpad, and its rate within
Project~A moves from $2.7\%$ to $17.7\%$ across a silent month we located by inspecting the
series rather than specifying in advance. Why, we cannot say: the author may simply
have used an agent more after the boundary, and neither repository records a tool-settings
history or which process authored a commit (Section~\ref{sec:disclosure}). What survives
that uncertainty is the consequence for an auditor. A disclosure rate read off the commit
log is not an involvement rate, cannot be converted into one from repository evidence, and
is not stable enough within one project to compare against itself.

\paragraph{Contributions.}
\begin{enumerate}
\item A field-by-field taxonomy of what an ordinary Git repository can, cannot, and only
apparently records about the provenance of a change (Section~\ref{sec:taxonomy}),
separating fields that are unrecoverable from fields that are recoverable but misleading.
\item A legibility measurement over two AI-trace artifacts (Section~\ref{sec:retention}):
what share of the lines ever committed to each is visible at the current revision, which
revisions survive into it, and how each artifact compares to every other tracked text file
in its own repository.
\item A measurement of the commit log's disclosure channel (Section~\ref{sec:disclosure}):
its sensitivity against an explicit within-repository proxy for agent activity, and the
instability of its rate over time, with the alternative explanations we cannot rule out.
\item Requirements for any system intending to capture research provenance in situ
(Section~\ref{sec:requirements}), derived field by field from the taxonomy's gaps.
\end{enumerate}

Our analysis produces an auditable ledger behind every number: one row per artifact revision,
disclosing commit, tracked text file and blame cohort.

\section{Related work}

\paragraph{Provenance and the integrity of the record.} Formal provenance models specify
what a computational result must record to be accountable: entities, the activities that
produced them, and responsible agents \citep{moreau2011opm,freire2008provenance}. The reproducibility literature consistently finds that this
record is not a by-product of ordinary practice and has to be deliberately produced
\citep{peng2011reproducible,stodden2016enhancing,pineau2021improving}. Supply-chain work adds
cryptographic attestation and transparency logs on top of the repository
\citep{torres2019intoto,newman2022sigstore}, binding an identity to an artifact but
presupposing that the events worth attesting were identified first. We measure that prior step: which provenance
fields survive in the Git repository of an AI-assisted project.

\paragraph{Mining repositories.} Repository mining has long-standing cautions about
treating Git history as a faithful record of development: commits are curated, rebased,
squashed and back-dated, and repository metadata systematically misrepresents authorship
and effort \citep{bird2009promises,kalliamvakou2016depth,gousios2012ghtorrent}.
We take the cautions as the object of study rather than as a threat to mitigate, and ask what
they imply for provenance claims about AI involvement.
The commit history of an AI-assisted research repository has also been mined for reusable
agent practice \citep{wu2026skillmine}, and scientific code repositories have been turned into
environments that agents learn from \citep{geng2026scienceideturningworldsscientific}. That work reads the repository
for what an agent can reuse, whereas we read it for what it can attest.

\paragraph{AI disclosure and its verification.}
Venue policies now require authors to declare AI use \citep{nature2023tools}, and studies of
the resulting literature estimate rising LLM involvement from textual signal
\citep{liang2024mapping}.
Benchmarks of autonomous research find that agents still depend heavily on human
methodological guidance \citep{zhou2026asibenchdawnartificialsuperintelligence}.
Reasoning models trained with reinforcement learning can fail when the partial work they build
on contains misleading or corrupted steps, unless they are explicitly trained to detect and
repair such steps, for example through adversarial self-play
\citep{li2026learningrobustreasoningguided}. Scoring the end-to-end output of research agents
takes expert-written rubrics even under controlled conditions
\citep{xu2026researchclawbenchbenchmarkendtoendautonomous}. Research is therefore shared
between author and agent, and that division is what a disclosure has to describe.
Verification from the artifact is harder than it looks: generated
text is not reliably detectable post hoc \citep{sadasivan2023can}, watermarking requires
cooperation at generation time \citep{kirchenbauer2023watermark}, and integrity failures such
as fabricated references are recoverable only by checking the claim, not the process
\citep{walters2023fabrication,birhane2023science}. Closest to us, repository studies already
count AI self-declaration in the wild from commit messages and co-author trailers
\citep{kashif2025selfdeclaration,tufano2024unveiling,liu2026debt}. They take the declaration
as the measurement; we ask what it can establish.

\paragraph{Structured disclosure.} Datasheets and model cards established that useful
disclosure requires a fixed schema with named fields, including fields whose answer is ``not
recorded'' \citep{gebru2021datasheets,mitchell2019modelcards}, which
Section~\ref{sec:requirements} applies to process provenance.

\section{Method}
\label{sec:method}

\subsection{Unit of analysis}

The unit of analysis is the \emph{AI-trace artifact}: the file or directory an agentic tool
maintains in the repository to carry state between sessions. Taking the artifact rather than
the project as the unit is what makes the two repositories comparable, since their research
content is not, and the trace is what an auditor would read. Table~\ref{tab:arms} states the
two cases and its caption what they share, as a description rather than a design matrix.

\begin{table}[t]
\centering
\footnotesize
\caption{The two cases. Both are single-investigator projects by the same author in the
environmental sciences, in Git repositories on the same provider. Every row but the last differs, and the differences are not separable. Counts are taken at a fixed census commit per
project, 2026-08-21 and 2026-02-03.}
\label{tab:arms} \begin{tabular}{lll} \toprule & \textbf{Project A} & \textbf{Project B}
\\ \midrule Agentic tool & Claude Code, Cursor & Cursor \\
Trace class & prescriptive & narrative \\
Trace artifact & directory of standing rule files & single running scratchpad \\
Repository commits & $466$ & $103$ \\
Commits revising the trace & $15$ & $39$ \\
Trace active & 2026-04-06 to 2026-05-18 & 2025-10-02 to 2026-01-20 \\
Artifact git-ignored & no & no \\
\bottomrule
\end{tabular}
\end{table}

Project~B's AI involvement is stated exactly as we can evidence it and no further: it is a
Cursor-assisted project, evidenced by a versioned agent scratchpad tracked across $39$
commits. We do not claim a proportion of the code was model-written, because nothing in the
repository would support such a claim, and the inability to support it is part of our
result.

One premise deserves stating, because a later result rests on it. The file at
\texttt{.cursor/\allowbreak scratchpad.md} is a project convention, not a format the tool creates and
maintains on its own: it lives in the tool's configuration directory, holds plans, findings
and running status in the tool's working idiom, and the project kept it under version
control. We treat revisions to it as a proxy for contemporaneous agent activity, imperfect
in both directions: a human can edit the file directly, and an agent session can produce
work committed later, separately, or alongside unrelated changes. Every claim in Section~\ref{sec:disclosure} that rests on it is marked as such.

\subsection{Measurement}

Four scripts run over both repositories, a fifth turns their output into every coordinate
plotted here, and a sixth checks each number printed in this paper against them. Per case we walk every commit touching the trace artifact, over the
union of the paths it has occupied, and record per-revision insertions and deletions; count
the lines the census revision still shows; check whether the artifact is git-ignored;
extract commit-message trailers; and compare dates the artifact asserts about itself against
the commit that introduced each line. Three measurement decisions are worth
stating, because the naive alternatives give indefensible numbers.

First, \textbf{the query decides part of the answer.} Project~A's trace directory was
renamed partway through its life, and Git offers three defensible ways to ask one artifact
for its history. Table~\ref{tab:queries} runs all three at the same census, and they
disagree about different things in both projects. The history-only share barely moves,
$\qspreadA$ points across the three forms in Project~A and $\qspreadB$ in Project~B, so this
paper's headline does not rest on which query an analyst ran. The episode record does move:
Project~A's revision count varies by a factor of $\qratioA$ and its apparent start date by
$\qdaysA$ days, because a path-limited log cannot see a rename whose source that same limit
filters out. The earlier objects were never lost, and nothing in the narrow answer flags the
shortfall. We report the path-union form throughout, since a path-limited log is what an
audit would run.

\begin{table}[t]
\centering
\footnotesize
\begin{tabular}{@{}lccc@{}}
\toprule
How the history was asked for & Revisions & Earliest dated & History-only \\
\midrule
\multicolumn{4}{@{}l}{\emph{Project~A}, whose trace directory was renamed once} \\
\quad \texttt{log -{}- <path it holds now>} & \qsinglerevA & \qsinglefirstA & $\qsingleshareA\%$ \\
\quad \texttt{log -{}- <every path it has held>} & \qunionrevA & \qunionfirstA & $\qunionshareA\%$ \\
\quad \texttt{log -{}-follow}, one file at a time & \qfollowrevA & \qfollowfirstA & $\qfollowshareA\%$ \\
\addlinespace
\multicolumn{4}{@{}l}{\emph{Project~B}, whose scratchpad never moved} \\
\quad \texttt{log -{}- <path it holds now>} & \qsinglerevB & \qsinglefirstB & $\qsingleshareB\%$ \\
\quad \texttt{log -{}- <every path it has held>} & \qunionrevB & \qunionfirstB & $\qunionshareB\%$ \\
\quad \texttt{log -{}-follow}, one file at a time & \qfollowrevB & \qfollowfirstB & $\qfollowshareB\%$ \\
\bottomrule
\end{tabular}
\caption{The same artifact, the same census commit, three defensible history queries. The share a
checkout does not show is nearly stable across them; the number of revisions and the date the
record appears to begin are not. Project~B's artifact never moved, which is why its three
rows nearly agree and Project~A's do not.}
\label{tab:queries}
\end{table}

Second, \textbf{disclosure is counted from trailers, not from message text.} The same AI
vocabulary matched as a substring anywhere in a message reports $\kwrateA\%$ for Project~A
against the trailer's $14.2\%$, and $\kwnotrailerA$ of those $\kwhitsA$ commits carry no
\texttt{Co-Authored-By} at all. Among those the commonest match is
\texttt{\sweeptopwordA}, on $\sweeptopcountA$, of which $\keysecondA$ carry a second trailer
key that reaches the sweep as prose; \texttt{ai}, \texttt{domain} and \texttt{chains} follow. A sweep cannot tell a
structured claim from a word, so we read keys: the headline rate uses
\texttt{Co-Authored-By}, and Section~\ref{sec:disclosure} reports what that misses.

Third, \textbf{the census is closed.} Project~A is a live repository and this paper was
written inside it, so the commits that produced the paper carry an AI co-author trailer and
would raise the rate they report. We pin each case to a fixed census commit, for Project~A the
last predating this paper, and read surviving lines from it, so neither untracked files nor later edits can move
them.

\section{What a repository can and cannot recover}
\label{sec:taxonomy}

Table~\ref{tab:taxonomy} takes the provenance fields an AI-disclosure statement would need
and asks, for each, what a repository kept the way these two were kept provides.
\emph{Recoverable} means the field is present and trustworthy at face value within the
audited history; \emph{partial} means something related is present but does not answer the
question asked; \emph{absent} means nothing in the repository bears on it. A fourth category matters more:
fields that are present and \emph{misleading}, where reading at face value returns a
confident wrong answer.

The verdicts are about default practice, not about what Git makes possible: a project that
decides to can commit its prompts, hook messages against diffs, or keep intermediate states
on branches. None of that happened here, and none of it is what a project gets by using an
agentic tool and committing normally, which is the condition an auditor should expect.

\begin{table}[t]
\centering
\footnotesize
\caption{Provenance fields against what a normally-kept Git repository supplies. Verdicts
describe default practice, not the limit of what Git can be made to record. Fields marked
\textcolor{red}{misleading} are worse than absent: they return a confident answer that does
not mean what it appears to.}%
\label{tab:taxonomy}
\begin{tabular}{L{2.75cm}L{6.15cm}L{3.77cm}}
\toprule
\textbf{Field} & \textbf{Best available evidence in the repository} & \textbf{Verdict} \\
\midrule
Committer identity & \texttt{author} and \texttt{committer} name and email & Recoverable,
self-asserted \\
Change timing & author and commit timestamps & Recoverable, self-asserted \\
Self-asserted dates & artifact's own date vs. its commit date & Recoverable, checkable \\
AI involvement & \texttt{Co-Authored-By}, when written & \textcolor{red}{Misleading}
(\S\ref{sec:disclosure}) \\
Deleted content & absent from the working tree, reachable by walking the file's history &
\textcolor{red}{Misleading} at a glance, recoverable if sought (\S\ref{sec:retention}) \\
Model identity & free text inside the same trailer & Partial, tool-dependent \\
Prompt/instruction & standing rule files, if kept & Partial, prescriptive only \\
Session narrative & scratchpad file, if tracked & Partial, overwrite-prone \\
Accepted unchanged & not distinguishable from authored text & Absent \\
Intermediate states & discarded between commits & Absent \\
Context supplied & no record of what the model was shown & Absent \\
Message fidelity & free text, never checked against the diff & \textcolor{red}{Misleading} \\
History rewriting & local reflog only; force-pushes unlogged & Absent afterwards \\
\bottomrule
\end{tabular}
\end{table}

Three rows carry the argument. \emph{Accepted unchanged} is absent: the questions that
separate assistance from authorship all reduce to whether a human exercised judgement over
model output, and a repository records only the state after the human decided, storing an
accepted suggestion and a hand-written line identically. \emph{AI involvement} is present
but misleading, quantified in Section~\ref{sec:disclosure}. \emph{Self-asserted dates} is
the one row where the repository does better than the artifact, because Git records when
each line was introduced, so a date an artifact asserts about itself can be checked against
the commit that wrote it. Run over the current text of both artifacts, that check returned
exactly one disagreement: a section header reading \texttt{2025-01-XX} on a line written on
2026-01-20, an unfilled template placeholder rather than a false claim about a date. We
report it to establish that the cross-check is implementable and that nothing runs it, not to
characterize how often self-dating fails.

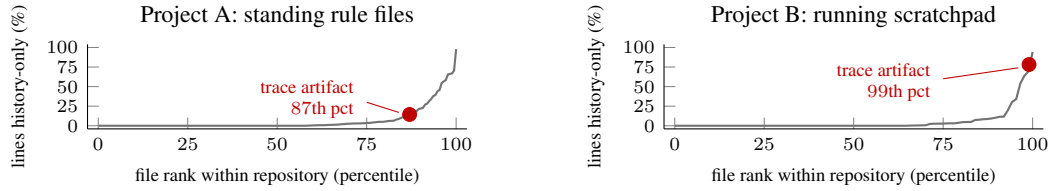
\begin{figure}[t]
\centering
\pgfplotsset{baseax/.style={width=6.6cm, height=2.7cm, ymin=-4, ymax=104,
    xmin=-3, xmax=103,
    xlabel={file rank within repository (percentile)},
    ylabel={lines history-only (\%)},
    tick label style={font=\scriptsize}, label style={font=\scriptsize},
    axis lines*=left, ytick={0,25,50,75,100}, xtick={0,25,50,75,100}}}
\begin{tikzpicture}
\begin{axis}[baseax, title={\small Project A: standing rule files},
             title style={font=\small, yshift=-3pt}, name=pa]
\addplot[black!55, thick] coordinates {\baselinecurveA};
\addplot[only marks, mark=*, mark size=2.6pt, red!75!black]
  coordinates {\baselinemarkA};
\node[font=\scriptsize, anchor=east, red!75!black, align=right]
  at (axis cs:74,34) {trace artifact\\ $\baselinepctA$th pct};
\draw[red!75!black, thin] (axis cs:75,30) -- (axis cs:84,17);
\end{axis}
\begin{axis}[baseax, title={\small Project B: running scratchpad},
             title style={font=\small, yshift=-3pt},
             at={(pa.right of south east)}, anchor=left of south west, xshift=13mm]
\addplot[black!55, thick] coordinates {\baselinecurveB};
\addplot[only marks, mark=*, mark size=2.6pt, red!75!black]
  coordinates {\baselinemarkB};
\node[font=\scriptsize, anchor=east, red!75!black, align=right]
  at (axis cs:74,56) {trace artifact\\ $\baselinepctB$th pct};
\draw[red!75!black, thin] (axis cs:75,60) -- (axis cs:96,74);
\end{axis}
\end{tikzpicture}
\caption{Every other tracked text file in each repository with at least $20$ lines ever inserted,
measured the same way as the trace artifact and sorted by history-only share. The $20$-line
floor limits small-denominator instability. Both medians
are $0\%$: most files in both repositories are fully visible, and the trace artifacts are
not. The comparison is internal to each repository, so it does not depend on
the two projects being alike.}
\label{fig:baseline}
\end{figure}

\section{Legibility: what the current revision shows}
\label{sec:retention}

We measure one quantity per artifact and state it precisely, because the natural paraphrases
of it are all wrong. For each artifact we sum the insertions Git attributes to it across
every commit that touches any path it has occupied, and compare that total against the
lines present at the census revision. The denominator is gross committed insertions, so a line
rewritten three times counts three times; it is not a count of distinct sentences or
episodes. Text overwritten within a session and never committed is invisible here, as is
anything on an unmerged branch.

The prescriptive
trace accumulates: $2{,}236$ lines committed, $321$ later removed, $1{,}915$ present now, so
$14.4\%$ of what was committed is no longer at the current revision. The narrative trace
churns: $1{,}537$ committed, $1{,}202$ removed, $335$ present, $78.2\%$ history-only. Across
$39$ revisions it grew by a net $335$ lines while absorbing $2{,}739$ lines of change, the
signature of a document rewritten in place rather than extended. Neither figure is a
storage failure. Git kept all $1{,}202$ lines and hands them to anyone who asks the file for
its history; it does not show them to anyone who asks for its contents.

\paragraph{Read against the repository's own files.}
A history-only share is uninterpretable on its own: a reader cannot tell whether $78.2\%$ is
remarkable or simply what files do. We therefore measured every other tracked text file in
each repository the same way and placed the artifact in its own repository's distribution
(Figure~\ref{fig:baseline}). Both distributions concentrate at zero: the median file is fully
visible at the census revision, and $58.0\%$ of Project~A's files and $67.0\%$ of
Project~B's have no history-only content at all. Against that, Project~A's rule directory
sits at the $87$th percentile of $174$ files, with $22$ above it, and Project~B's scratchpad
at the $99$th of $88$, with exactly one file above it: the agent's notebook is the second
most overwritten text file in the repository that contains it. Whatever separates the two
projects, each trace artifact is unusual by the standard of the files sitting next to it.

\begin{figure}[t]
\centering
\pgfplotsset{survax/.style={width=6.6cm, height=2.6cm, ymin=0, xmin=0,
    ybar, bar width=1.3pt, bar shift=0pt, area legend,
    xlabel={revision of the artifact, oldest to newest},
    ylabel={lines}, ylabel style={yshift=-4pt},
    tick label style={font=\scriptsize}, label style={font=\scriptsize},
    axis lines*=left, legend style={font=\scriptsize, draw=none, at={(0.03,0.97)},
      anchor=north west, fill=none}}}
\begin{tikzpicture}
\begin{axis}[survax, title={\small Project A: \survpctA\% still on screen},
             title style={font=\small, yshift=-3pt},
             name=sa, xmax=\survnA, ymax=930]
\addplot[fill=black!22, draw=none] coordinates {\survaddA};
\addplot[fill=blue!62, draw=none] coordinates {\survaliveA};
\legend{lines written, still visible}
\end{axis}
\begin{axis}[survax, title={\small Project B: \survpctB\% still on screen},
             title style={font=\small, yshift=-3pt},
             at={(sa.right of south east)}, anchor=left of south west, xshift=13mm,
             xmax=\survnB, ymax=250]
\addplot[fill=black!22, draw=none] coordinates {\survaddB};
\addplot[fill=blue!62, draw=none] coordinates {\survaliveB};
\end{axis}
\end{tikzpicture}
\caption{What each revision wrote (grey) against how much of it a reader of the census
revision still sees (blue), by \texttt{git blame} at that revision, so a line counts only if
its text is unchanged. Both panels count $41$ blame cohorts, not the $15$ and $39$ revisions
the path query returns, because cohorts follow the file across renames. Panel percentages are shares of those totals, so they sit just below the text's.}
\label{fig:survival}
\end{figure}
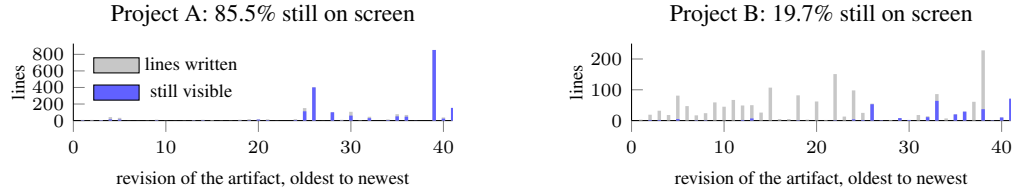

\paragraph{Which revisions a reader can still see.}
An aggregate share says how much of the record is off the page, not which part. We asked
\texttt{git blame} at the census revision which commit last wrote each visible line, and
grouped the answers by revision (Figure~\ref{fig:survival}). Counted that way the two
artifacts differ far less than $85.6\%$ against $21.8\%$ suggests. $\survzeroA$ of Project~A's
$\survnA$ revisions leave no surviving line at all, against $\survzeroB$ of Project~B's
$\survnB$, so in both an auditor reading the file learns nothing about those revisions beyond
that they happened. Both are top-heavy, Project~A the more so: its newest $\survhalfA$
revisions, $\survhalfpctA\%$ of the history, supply half of everything visible, against
Project~B's newest $\survhalfB$ at $\survhalfpctB\%$. Project~A's line share is high because
one revision contributes $\survmaxA$ of the $\survvisiblefmtA$ visible lines, not because its
history survives evenly. Either artifact, opened at its current revision, is closer to a
recent snapshot than to a record of the project, and gives no indication of which it is.

\paragraph{The obvious query understates the artifact's age.}
The directory is not where its contents were written. Its first thirteen files entered it by
rename on \survpathfirstA, from a rules directory elsewhere in the same repository, and more
moved in from two other locations that week. A path-limited log therefore reports
those lines as insertions on the day they moved. Blame follows a file across renames, and
attributes $\survprefmtA$ of the $\survvisiblefmtA$ lines a reader sees, $\survprepctA\%$, to
$\survprenA$ commits the path query does not list, the earliest \survfirstA. Both queries are
correct, but the one an auditor would run dates content to the day it moved, so the artifact
reads as seven months younger than it is.

\paragraph{Visibility is not the same as usefulness.}
The prescriptive trace keeps most of its lines on screen, but the lines are rules:
statements of what should be done in future. A rule records the conclusion drawn from an
episode and discards the episode. Section~\ref{sec:reconstruction} reports what happened
when we tried to recover why any given rule exists: the artifact supplied the rule's text
and its commit date and nothing about the failure that motivated it. We make no
quantitative claim about information content. The two shortfalls are complementary
rather than one case outperforming the other.

\paragraph{Both cases depend on a non-default choice.}
Neither artifact is git-ignored, which we verified rather than assumed, and nothing required
it: an agent's working directory is ordinary local scratch state, and excluding it is a
one-line edit neither tool prompts for and neither warns against. Had either author made
that edit before first committing the artifact, every number here would be zero.

\section{Disclosure: what the commit log says}
\label{sec:disclosure}

Git trailers are arbitrary key-value lines and Git fixes no key set;
\texttt{Co-Authored-By} became the AI field by convention, because the best-known agentic
tools write it, so it is what an automated audit keys on. Project~A carries one naming an AI
identity on $66$ of $466$ commits ($14.2\%$), Project~B on $1$ of $103$ ($1.0\%$). Read
naively, those rates say Project~A involved roughly fourteen times as much AI as Project~B.
Three measurements show that reading cannot be sustained.

\paragraph{The audited key is not the only key.}
An audit's choice of key is therefore part of its result. Scanning
Project~A's trailer blocks without fixing a key in advance finds $\keynA$ whose value names
an AI system: \texttt{Co-Authored-By} on $66$ commits, \texttt{Made-with} on $\keysecondA$ naming the other
tool, and \texttt{Claude-Session} on $\keythirdA$. No commit carries both \texttt{Made-with} and
\texttt{Co-Authored-By}, and \texttt{Claude-Session} appears only beside the latter, so the
keys divide the record rather than corroborate it: $\keyunionA$ of $466$ commits
($\keyunionpctA\%$) carry one. An audit reading \texttt{Co-Authored-By} misses $\keymissedA$,
$\keymissedpctA\%$ of those the repository holds, and both measurements below move with the choice: the
cross-tabulation rises from $5$ of $15$ to $\keytraceA$, the temporal step from a factor of
$6.6$ to $\keyerafactorA$. Project~B carries one key and its two rates coincide, so the
shortfall is a property of the repository read, not a constant to correct for.

\paragraph{The trailer has low sensitivity against the repository's own proxy.}
Cross-tabulate the trailer against the scratchpad. In Project~B, $39$ commits revise the
file the project maintained to hold the agent's working notes, and $1$ ($2.6\%$) carries an
AI trailer; in Project~A the same cross-tabulation over the rule files gives $5$ of $15$
($33.3\%$). Conditional on that proxy, the trailer's sensitivity to
contemporaneous agent activity is neither close to the repository-level rate nor consistent
across cases.

Project~B's single disclosing commit edits the same scratchpad the other $38$ revisions
edit, and names a different agentic tool from the one the project otherwise used. Like every
field in the log it is self-asserted, so we report it as description, not identification.
What it establishes is that two commits on one file, by one author, can differ
entirely in whether they disclose.

\paragraph{The rate is unstable within a single project.}
\begin{figure}[t]
\centering
\begin{tikzpicture}
\begin{axis}[
  ybar stacked,
  width=13.3cm, height=3.2cm, ymin=0, ymax=280,
  ytick={0,100,200},
  ylabel={commits in month},
  symbolic x coords/.expand once=\monthcoordsA,
  xtick=data,
  tick label style={font=\small}, label style={font=\small},
  xticklabels/.expand once=\monthlabelsA,
  x tick label style={align=center, font=\tiny},
  bar width=13pt, clip=false, area legend,
  legend style={at={(0.015,0.98)}, anchor=north west, draw=none,
                font=\scriptsize, fill=none, row sep=-1pt},
  axis lines*=left, enlarge x limits=0.045,
]
\addplot[fill=blue!62, draw=none] coordinates {\monthdiscA};
\addplot[fill=black!18, draw=none] coordinates {\monthrestA};
\legend{commits disclosing AI, commits not disclosing}
\draw[dashed, black!55, thick]
  ([xshift=-15pt]axis cs:2026-02,0) -- ([xshift=-15pt]axis cs:2026-02,275);
\node[font=\scriptsize, anchor=north east, align=center]
  at ([xshift=-19pt]axis cs:2026-02,272) {$3/111$ commits\\ $\mathbf{2.7\%}$};
\node[font=\scriptsize, anchor=north west, align=center]
  at ([xshift=-11pt]axis cs:2026-02,272) {$63/355$ commits\\ $\mathbf{17.7\%}$};
\end{axis}
\end{tikzpicture}
\caption{Project~A, monthly commit volume split by whether the commit carries an AI
co-author trailer. The repository has no commits in 2026-01, so the boundary coincides with
a gap in activity, and we located it by inspecting the series rather than specifying it in
advance.}
\label{fig:disclosure}
\end{figure}
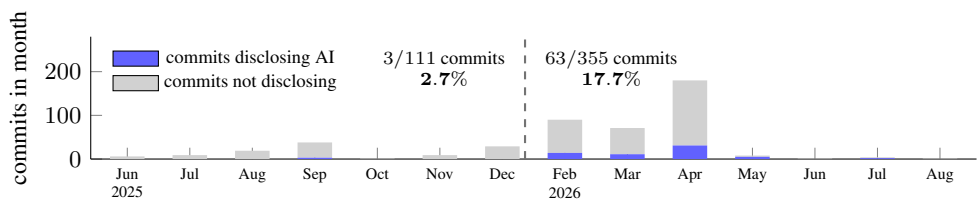

Figure~\ref{fig:disclosure} splits Project~A's commits by month. Disclosure is not spread
evenly over the history: it is near-absent through 2025-12, at $3$ of $111$ commits
($2.7\%$), and sustained afterwards, at $63$ of $355$ ($17.7\%$), a factor of $6.6$. Two
caveats belong with that number. We chose the boundary after seeing the monthly series, so
$6.6$ is the largest contrast available rather than a test of a prior hypothesis. And the
repository is silent through 2026-01, so this is a resumption rather than a clean
adjacent-month change, and whatever else changed in the gap changed with it.

\begin{figure}[t]
\centering
\begin{tikzpicture}
\begin{axis}[
      xbar, scale only axis, width=8.2cm, height=2.4cm, xmin=0,
  ytick/.expand once=\identticksA,
  yticklabels/.expand once=\identlabelsA,
  tick label style={font=\scriptsize},
  bar width=7pt, axis y line*=left, axis x line=none, ytick style={draw=none},
  enlarge y limits=0.11, enlarge x limits={upper, value=0.09},
  point meta=x, nodes near coords, nodes near coords style={font=\scriptsize},
]
\addplot[fill=blue!52, draw=none] coordinates {\identcountsA};
\end{axis}
\end{tikzpicture}
\caption{The identity string carried by Project~A's $\identtotalA$ AI trailers. Five values
name a model version and one is a bare vendor name; two of the five differ only by a
context-window suffix and denote the same model.}
\label{fig:identity}
\end{figure}
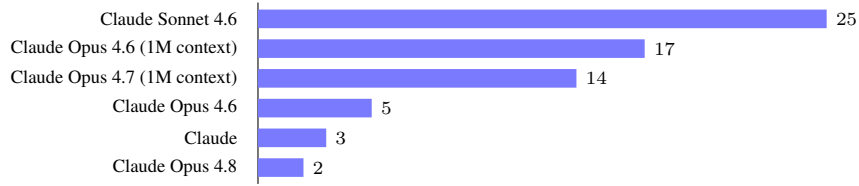

The trailer text is the only further evidence the log offers, and it is free text. Across Project~A's $\identtotalA$ disclosing trailers it takes $\identnA$ distinct values (Figure~\ref{fig:identity}), so a policy asking which model was used receives $\identnA$ answers for at most four models and $\identbareA$ commits answering nothing. The strings name which software was open, not how much the author relied on it.

\paragraph{What we can and cannot conclude.}
The rival explanation is obvious and we cannot dismiss it: the author may simply have used
an agent more, and asked it to write commits more often, after the gap. On that reading the
trailer is a noisy but positively informative signal, not an artifact, and it fits every
number we report. Rejecting it would need a record of the tool's settings over time or of
which process authored each commit, and neither repository contains either. Such settings do
move without notice: one widely used editor switched its AI co-author trailer on by default
in one release and reverted it two releases later, and is reconsidering the key name
\citep{vscode2026coauthor}. We flag this as
the section's central limitation and do not claim the mechanism.

What survives it is the part that matters for an audit. A disclosure rate read off the commit
log is a rate of commits that say so, and no repository evidence converts it into a rate of
commits a model helped write. It undercounts: a project evidenced as agent-assisted reports $1.0\%$. The undercount is not a constant to correct by either. It varies
with the key read, which costs $\keymissedpctA\%$ of Project~A's disclosures and none of
Project~B's; with the commit, since the trailer's sensitivity against a within-repository
proxy is $33.3\%$ in one case and $2.6\%$ in the other; and with the date, since the rate
moves by a factor of $6.6$ inside a single project. Pooling over a corpus has no reason to
average errors of that shape away. The field is \textcolor{red}{misleading} in the sense of
Table~\ref{tab:taxonomy}: it returns a confident number that no repository evidence turns
into the quantity a reader will take it for.

\section{Reconstruction from a prescriptive trace}
\label{sec:reconstruction}

We attempted the task an auditor would face: reconstruct, from Project~A's prescriptive
trace alone, the episodes that produced its rules. The trace holds
standing rules on citation, figure specification and assumption checking, each recoverable as
text with a commit date. We code each rule on three dimensions kept separate, because
collapsing origin, human response and evidential basis into a single label is what makes such
codings unfalsifiable: \emph{origin evidence} is AI, human, mixed or \texttt{unknown};
\emph{human review state} is accepted unchanged, modified or \texttt{not observed};
\emph{evidence strength} runs from direct transcript through tool metadata and
contemporaneous note down to \texttt{inference only}. The coding judges what evidence exists
for a claim, not what caused a change: we do not attempt to establish that a model rather
than a person originated any particular line, because Section~\ref{sec:taxonomy} shows the
repository cannot settle it. The pattern is uniform, and \texttt{unknown} is the result
rather than a gap in it. Origin evidence is \texttt{unknown} for every rule, since nothing
distinguishes a rule the author wrote from one the agent drafted and the author kept; review
state is \texttt{not observed} for the same reason; and evidence strength reaches
\texttt{tool metadata} at best, and \texttt{inference only} for why a rule exists.

We do not report whether a rule prevented recurrences: the absence of a corrective commit
is equally consistent with the problem being fixed silently, not occurring, or not being
detected. A prescriptive trace supports claims about what the project's
standards were at a given date; on who originated the work it offers only a self-asserted
author field and whatever the message says.

\section{Requirements for in-situ capture}
\label{sec:requirements}

Each requirement is derived from a gap measured above and stated tool-agnostically.

\begin{enumerate}
\item \textbf{Record the accept event, not just the result.} Whether a human modified model
output separates assistance from authorship, and a repository observes only post-decision
state. Capture has to happen at the moment of acceptance, in the editor or the agent.

\item \textbf{Separate the involvement field from the tool-identity field, and name both.} A
trailer answers ``which tool wrote this message'' and is read as ``how much did a model
contribute''. These need different fields under keys no tool is free to invent, since
$\keynA$ competing keys in one repository are $\keynA$ fields an audit will not read
together; and the involvement field needs a defined value for ``not recorded'' so silence is
not read as absence \citep{gebru2021datasheets}.

\item \textbf{Present the narrative, do not merely store it.} Git holds all $78.2\%$ absent
from the current revision, so this is not a storage failure: a mutable document shows its
latest state, and reading its history is a step nobody is prompted to take. Records should be
append-only, or exposed as a series rather than a head.

\item \textbf{Make the tracking decision explicit.} Every legibility figure here is
contingent on the artifact not being git-ignored, a choice made once and invisibly. A capture
system should surface, when the artifact is created, that excluding it discards the trace.

\item \textbf{Bind records to history rather than to paths.} A path query without rename
following returned less than half of one artifact's revisions, with no sign it was
partial. Provenance queries must follow renames and date content by when it was written, not when it
last moved.

\item \textbf{Check self-asserted dates against recorded ones.} An artifact that dates itself
can disagree with the history containing it, as one does here, and the disagreement is
mechanically detectable from data the repository already holds. Neither tool ran this check.
\end{enumerate}

\section{Threats to validity}
\label{sec:threats}

\paragraph{Two repositories, one author, and confounded conditions.}
Holding the author constant removes between-person variation
and nothing else: the projects differ in tool, trace format, subject,
duration and year, all aligned with the outcome. We attribute no contrast to the trace format, and do not
claim Section~\ref{sec:requirements} is complete on two repositories. What the design supports is existence: a repository can hold a large record and
present a small one, and a disclosure field can vary sixfold within one author's history.

\paragraph{Single coder, outcome-aware.}
The coding in Section~\ref{sec:reconstruction} was done by an author
who created both repositories and knew the results. We cannot report
inter-rater agreement \citep{cohen1960coefficient}, and do not present the coding as an
independent measurement. Two things limit the exposure: the quantitative results, which carry
the paper, are mechanical and regenerable; and the coding judges evidence availability rather than causation, so its outcomes, \texttt{unknown}
and \texttt{not observed}, are the ones a motivated coder had least reason to assign. That is an argument, not a control.

\paragraph{The answer depends on the query.}
Table~\ref{tab:queries} measures this: over three defensible query forms the history-only share moves at most $\qspreadA$ points, while the revision count
and the apparent start date do not hold still. Headline figures are the path query's, what the
obvious audit returns. Cohort survival comes from \texttt{blame}, which
follows renames and does not simplify history, so it counts
$\survnB$ revisions and $\survaddedfmtB$ lines for Project~B where the path query sees $39$
and $1{,}537$.

\paragraph{Measurement scope.}
The history-only share is measured in lines, which weights a rewritten paragraph
and a deleted heading equally. We measure the trace artifact only: AI contributions elsewhere are
outside it and unidentified in the log, so we claim nothing about their size.

\section{Ethics}

The material is two repositories authored by one of us; no third party's data is involved. Commit metadata nonetheless carries personal information, so identifying strings, addresses, repository names and paths are excluded from the paper and
the ledger, which replaces subject lines with a conventional-commit type and keys rows by
a salted hash of the commit id. Model identity strings are reported because they name
software, not people. Anonymization is procedural: a reader who knows the field could locate these repositories from
the counts, dates and tool names, which are the measurement.

We disclose our own AI use in the terms we argue for: text, analysis scripts and figures were
produced with an agentic assistant under author review, inside Project~A, so our own commits
carry the trailers Section~\ref{sec:disclosure} audits.

\section{Conclusion}

Asked to disclose how AI contributed to a piece of research, an author with a well-kept
repository has less evidence than policy assumes. One artifact records the process and shows a
fifth of it; the other stays visible and records conclusions; and the field an audit reads
misses $\keymissedpctA\%$ of one repository's disclosures. Until these facts are captured as
they happen, a repository can constrain a provenance claim but not establish one.

\begin{ack}
We thank Zong-Liang Yang and Chenxi Hu for discussion of the two projects.
We acknowledge high-performance computing support from the Derecho
(doi:10.5065/qx9a-pg09) and Casper (\url{https://ncar.pub/casper}) systems provided by the
NSF National Center for Atmospheric Research (NCAR), sponsored by the National Science
Foundation. We also acknowledge the Texas Advanced Computing Center (TACC) at The University
of Texas at Austin for providing computational resources that have contributed to the
research results reported within this paper.
\end{ack}

\bibliographystyle{plainnat}
\bibliography{references}

\end{document}